\documentclass[showpacs,preprint,aps]{revtex4}
\usepackage{graphicx}
\usepackage{dcolumn}
\usepackage{bm}
\usepackage{amssymb}
\usepackage{amsmath}

 \newenvironment{Proof.}[1][Proof.]{\begin{trivlist}
     \item[\hskip \labelsep {\bfseries #1}]}{\end{trivlist}}
\begin{document}
\setcounter{page}{1}
\title
{On Born approximation in black hole scattering}
\author
{D. Batic$^{1}$, N. G. Kelkar$^2$ 
and M. Nowakowski$^2$} \affiliation{ $^1$
Department of Mathematics, University of West
Indies, Kingston 6, Jamaica\\
$^2$ Departamento de Fisica, Universidad de los Andes, Cra.1E
No.18A-10, Bogota, Colombia}

\begin{abstract}
A massless field propagating on spherically
symmetric black hole metrics such as the Schwarzschild,
Reissner-Nordstr\"{o}m and Reissner-Nordstr\"{o}m-de Sitter
backgrounds is considered. In particular, 
explicit formulae in
terms of transcendental functions for the scattering of massless
scalar particles off black holes are derived within a Born approximation. 
It is shown that 
the conditions on the existence of the Born integral  
forbid a straightforward extraction of the quasi normal modes using 
the Born approximation for the scattering amplitude.
Such a method has been used in literature. 
We suggest a novel, well defined method, to extract the large imaginary 
part of quasinormal modes via the Coulomb-like phase shift. 
Furthermore,
we compare the numerically evaluated exact scattering amplitude 
with the Born one to find that the approximation is not very useful 
for the scattering of massless scalar, 
electromagnetic as well as gravitational 
waves from black holes.
\end{abstract}
\maketitle

\section{Introduction}
The theory of the scattering of massless and massive particles
with different spins off black holes 
has a long history in gravity and quantum gravity 
\cite{BHscattering}.
Over the years the topic attracted much attention and led to
the discovery of new phenomena like glory scattering \cite{glory}, 
superradiant scattering and
black hole bomb \cite{BHbomb}. 
A related subject is   
the modes of perturbation 
of black holes (characteristic sound of a black hole), the so-called Quasi
Normal Modes (QNMs) \cite{QNM}, which can be also extracted 
from the corresponding scattering $S$-matrix where they appear as complex poles
(resonances). The relevance of QNMs for the quantum gravity lies, e.g.,  
in a method to quantize the area (and therefore also the entropy) of
a black hole.
However, the ramification of the $S$-matrix poles goes beyond the
QNMs as modes of perturbation. The poles normally harbour all the information
about the existence of stable bound states, quasi-bound states and
resonances \cite{Smatrix}. For instance, the absence of a stable bound
state of a particle in a black hole metric \cite{us} must
get reflected in the properties of the poles of the $S$-matrix which
in such a case should always have a non-zero imaginary part. 
For massless fields the scattering amplitude has, in addition, a cut
in the complex frequency plane beginning from the zero frequency.
The lack of bound state of a spin-$1/2$ particle in certain
black hole spacetimes \cite{us} indicates a non-local
quantum effect in semi-classical black hole physics since 
the classical picture yields a stable orbit.
It is therefore not surprising that different
mathematical methods have been applied to the theory of black hole
scattering to probe into the semi-classical aspects of quantum gravity
\cite{Leaver, WKB, Cho}.
Recently a suggestion to extract QNMs from the Born
approximation to the scattering amplitude was made in \cite{CHO, PAD, VIS}
with apparently reasonable results for the imaginary part of the QNMs.
It is often tempting to use such simplified approximations to obtain
analytical results. In \cite{we3prd} it was shown that the extraction of QNMs 
from the Born amplitude for the Schwarzschild and Schwarzschild de Sitter case 
is mathematically not well founded. 
In the second reference in \cite{we3prd}, a regularization method to 
evaluate the otherwise divergent integral was suggested. However, the fact 
remains that the Born integral simply does not exist for a certain range 
of the energy parameter and it does not make sense to insist on extracting 
poles from such an integral. 
Here we will demonstrate that this feature is more general and that 
the Born integrals for all interesting spacetimes, 
including Reissner-Nordstr\"om and Reissner-Nordstr\"om de Sitter 
exist only within a restricted validity
range of the energy $k \equiv \omega$ which lies outside
the range of the extracted values of the QNMs. This is in agreement
with the observation that there are no $S$-matrix poles in the
Born approximation of the Feynman-Dyson perturbative expansion.
In an attempt to remedy this problem, we prove that the large imaginary parts  
of the QNMs can be obtained in a mathematically rigorous way through the 
Coulomb-like scattering phase. 
Secondly, it appears timely to make an explicit check of 
the limitations of the Born approximation in the physical
region of the phase space, i.e., its usefulness to describe
correctly the scattering process. Some work in this direction has been
already done in \cite{Matzner1, Sanchez1}. 
However, either the distorted wave Born approximation is used
\cite{Sanchez1} or the comparison with the Born approximation
is done for low frequencies and in terms of the differential
cross section $d\sigma/d\Omega$ paying attention to
the comparison using scattering angles. We perform 
a test of the validity of the Born approximation for different $l$ 
and in the case of the scattering of 
massless fields with spin $s$=0, 1, 2.
There also exists an ambiguity in the scattering 
phase shift in the Born approximation which appears in 
literature \cite{Dolan}. Hence there arises a general question 
whether it is reasonable to apply the Born approximation
for the scattering off black holes. We therefore make a
comparison between exact albeit numerical results 
obtained via a Riccati equation for the reflection 
coefficient ${\cal R}$ (in the context of a one dimensional
tunneling problem into which we can map the scattering problem) and
the Born approximation. The comparison shows that only for the spin $0$ case
and the $l=0$ partial wave, the agreement between the exact result and the
Born approximation is reasonable over a wide range of $\omega$. 
For spins, $s=1,2$, there is no agreement for any $l$. 

It is well known from general scattering theory \cite{Taylor} that
the Born approximation might work at high energies \cite{Taylor}.
However, if the required energies are so high that the 
reflection coefficient goes to zero, the
comparison might lose any physical meaning.
There exist circumstances where the Born approximation 
is expected to be bad and then again
circumstances where it is valid for some particular angular
momentum $l$ provided we mean the Born approximation
for a partial wave \cite{Taylor}. 
This is what happens at least in the case of $s=0$.

The article is organized as follows. We briefly outline the
formalism in section two. Section three is about the 
possibility to extract the QNMs from the Born amplitude.
In section four we obtain an alternative way to obtain the QNMs.  
In section five we compare the exact scattering amplitude results with those 
in the Born approximation. In section six we draw our conclusions.  

\section{Black hole scattering and the Born amplitude}
We consider the propagation of a massless scalar field
$\phi=\phi(t,r,\vartheta,\varphi)$ governed by the wave equation
$g^{\mu\nu}\nabla_\mu\nabla_\nu\phi=0$ where $g_{\mu\nu}$ denotes
a static, spherically symmetric black hole metric whose line
element is
\[
ds^2=f(r)dt^2-\frac{dr^2}{f(r)}-r^2(d\vartheta^2+\sin^2{\vartheta}d\varphi^2)
.
\]
Taking into account that the eigenvalues of the spherical Laplacian 
are $-\ell(\ell+1)$ the wave equation can be separated by means of
the ansatz
\begin{equation}\label{ansatz}
\phi(t,r,\vartheta,\varphi)=e^{i\omega t}\frac{1}{r}~\psi_{n \ell
\omega}(r)Y_{\ell m}(\vartheta,\varphi),\quad {\rm{Re}(\omega)>0}
\end{equation}
giving rise to the Schr\"{o}dinger-like equation for the radial
part
\begin{equation}\label{radial}
\left[-\frac{d^2}{dr_{*}^2}+V(r)\right]\psi_{n \ell \omega
}=\omega^2\psi_{n \ell \omega },
\end{equation}
where $V(r)=f(r)U(r)$ (with $f(r)$ having a different form for the 
various metrics considered in the next section) and 
\begin{equation}
U(r)=\frac{\ell(\ell+1)}{r^2}+\frac{f^{'}(r)}{r}.
\end{equation}
Here, a prime denotes differentiation with respect to $r$ whereas
$r_{*}$ is a tortoise coordinate defined through
\[
\frac{dr_{*}}{dr}=f(r)^{-1}.
\]
We recall that for static, spherically symmetric black hole
metrics which are asymptotically flat at spatial infinity the
quasinormal modes boundary condition requires that there is a
purely ingoing plane wave at the event horizon $r_0$ and a purely
outgoing plane wave at space-like infinity, i.e.
\begin{equation}\label{QNMs condition 1}
\psi_{n \ell \omega }(r)\propto\left\{ \begin{array}{ll}
         e^{i\omega r_{*}} & \mbox{if $r_{*}\to-\infty$},\\
         e^{-i\omega r_{*}} & \mbox{if $r_{*}\to+\infty$},\end{array} \right.
\end{equation}
whereas for black hole metrics which are asymptotically de Sitter
at spatial infinity the appropriate boundary condition for
analyzing quasinormal modes is that the wave function should be
outgoing at the cosmological horizon $r_c$ but vanishing at the
origin \cite{CHO}
\begin{equation}\label{QNMs condition 2}
\psi_{n \ell \omega }(r)\propto\left\{ \begin{array}{ll}
         0 & \mbox{if $r\to 0$},\\
         e^{-i\omega r_{*}} & \mbox{if $r\to r_c$}.\end{array} \right.
\end{equation}
Quasinormal modes can be found by computing the poles of the
scattering amplitude $S(\omega)$. To see that we write 
the boundary condition at $r_* \to \infty$ or $r \to r_c$ in the form
\begin{equation} \label{Sboundary}
A(\omega) e^{+i\omega r_*} + B(\omega)e^{-i\omega r_*}
\propto \left( e^{+i\omega r_*} + S(\omega)e^{-i\omega r_*}\right)
\end{equation}
where $S(\omega)$ is a component of the scattering matrix
${\bf S}$. Hence imposing $A(\omega)=0$ is equivalent
to looking for a subset of $S$-matrix poles. Since the problem
of black hole scattering is mapped into one dimension
it makes sense to make some identifications. 
The one dimensional problem is equivalent to a quantum
mechanical tunneling problem such that the reflection
amplitude $R(\omega)$ is simply identical to $S(\omega)$. 
In one dimension (1D) $S(\omega)$ is thus identified with
the scattering amplitude $f^{\rm 1D}(\omega)$ via
$S(\omega)= R(\omega) = if^{\rm 1D}(\omega)/2\omega$ \cite{Adhikari}. 
This is different from the identification in three dimensions (3D) 
where $S(\omega)=1+2i \omega f^{\rm 3D}(\omega)$.
In literature, while evaluating cross sections, considering the 
asymptotic behaviour of the wave function in the 3D radial coordinate $r$, 
generally $R(\omega)$ is identified with 
$S(\omega)= e^{2i\delta (\omega)}$ \cite{prldolan} 
where $\delta (\omega)$ is the scattering phase shift. 
The reflection and transmission in 1D 
can be identified as two channels in a scattering 
problem such that the $R(\omega)$ and 
$T(\omega)$ enter the elastic and absorption cross sections respectively.
The unitarity of the $S$-matrix, ${\bf S}^{\dagger}{\bf S}=1$, is then 
given by $|R(\omega)|^2 +|T(\omega)|^2=1$.
We shall discuss this in more detail in section IV.

In one dimension the Born approximation for the scattering amplitude 
is given by \cite{CHO}
\begin{equation}\label{born1D}
f_{Born}^{1D}(\omega)=\int_{-\infty}^{+\infty}dr_{*}~V(r_{*})e^{2i\omega
r_{*}}.
\end{equation}
In particular, for static, spherically symmetric black hole
metrics which are asymptotically flat, the scattering amplitude
reads
\begin{equation}\label{scattering amplitude 1}
f_{Born}^{1D}(\omega)=\int_{r_0}^{+\infty}dr~U(r)e^{2i\omega r_{*}(r)},
\end{equation}
whereas for those black hole backgrounds going over asymptotically
at spatial infinity to a de Sitter geometry we have
\begin{equation}\label{scattering amplitude 2}
f_{Born}^{1D}(\omega)=\int_{r_0}^{r_c}dr~U(r)e^{2i\omega r_{*}(r)}.
\end{equation}

Some comments are now in order. A first remark refers to the nature
of the two channel process which indicates that the phase shift
$\delta (\omega)$  cannot be real in general. Indeed, we have
$S(\omega)=e^{2i\delta (\omega)}=e^{-2{\rm Im} \delta (\omega)} 
e^{2i{\rm Re}\delta (\omega)}=
\eta e^{2i{\rm Re}\delta (\omega)}$ where $\eta$ is the inelasticity factor. 
If the reflection amplitude is identified with $S(\omega)$, any
usage of an approximation where the phase shift
comes out real necessarily neglects the transmission. For instance,
this applies to a formula like Eq. (35) 
for the Born phase shift mentioned in \cite{Dolan}.
A second remark concerns the difference between the Born approximation
in one and three dimensions. In three dimensions with a potential
depending only on $r$, the Born integral $\int e^{-i{\bf q}\cdot {\bf r}}
V(r) d{\bf r}$ is equal to $(-4\pi /q)\int_0^{\infty} (\sin(qr)/r)V(r) r^2dr$
and is hence real. The Born amplitude in (\ref{born1D}) 
is complex for a real $V(r)$. This implies that the conclusion 
that the phase shift in the Born approximation must be small \cite{Taylor} 
is not stringent here.
\section{Quasinormal Modes in the Born approximation}
In this section we discuss the Schwarzschild and Schwarzschild-de 
Sitter cases briefly and concentrate on the existence of quasinormal modes 
(QNMs) in the Reissner-Nordstr\"{o}m
and Reissner-Nordstr\"{o}m-de Sitter cases applying the Born approximation. 
\subsection{Remark on Schwarzschild and Schwarzschild-de Sitter metric}
In \cite{we3prd} it was demonstrated that the Born integrals have a certain 
range of validity for the above mentioned metrics. Indeed, for the 
Schwarzschild case, the Born integral is restricted to the following range: 
\begin{equation}\label{conditio}
0\leq\omega_I<\frac{1}{4M} \, ,
\end{equation}
where $\omega_I$ is the imaginary part of the quasinormal mode frequency 
$\omega=\omega_R+i\omega_I$, with the real part $\omega_R>0$. 
The Born integral in this case is a linear combination of products of the 
Gamma function and Whittaker function \cite{we3prd}. the latter possess the 
following poles: 
\begin{equation}\label{QNMSCH}
\omega_n=-in\kappa,\quad\omega_n=in\kappa \,,
\end{equation}
which, however, in view of the restrictions on the Born integral cannot be 
interpreted as quasinormal modes. The Schwarzschild-de Sitter case is 
similar and we refer the reader to \cite{we3prd} for further details. 

\subsection{Reissner-Nordstr\"{o}m metric}
In this case $f(r)=1-2M/r+Q^2/r^2$ where $M$ and $Q$ are the mass
and charge of the black hole respectively. We shall analyze the
scattering amplitude both for the non extreme $M>|Q|$ and extreme
case $M=|Q|$. In the non extreme case the black hole possesses two
distinct horizons at $r_{\pm}=M\pm\sqrt{M^2-Q^2}$ whose surface
gravity is given by the simple formula
\[
\kappa_{\pm}=\frac{r_{+}-r_{-}}{2r_{\pm}^2}.
\]
Notice that since $r_{+}>r_{-}$, 
$\kappa_{-}$ will always be greater than $\kappa_{+}$. 
The scattering amplitude is represented
by the integral
\[
f^{1D}_{Born}(\omega)=\int_{r_0}^{+\infty}dr~U(r)e^{2i\omega r_{*}(r)},\quad
U(r)=\frac{\ell(\ell+1)}{r^2}+\frac{2M}{r^3}-\frac{2Q^2}{r^4}.
\]
Taking into account that we are interested in the region $r>r_{+}$
the tortoise coordinate $r_{*}$ will be given by the following
expression \cite{MAT}
\[
r_{*}=r+\frac{1}{2\kappa_{+}}\ln{\left(\frac{r}{r_{+}}-1\right)}-\frac{1}{2\kappa_{-}}\ln{\left(\frac{r}{r_{-}}-1 \right)}.
\]
In view of the above relation it is not difficult to see that we
have to evaluate integrals of the form
\[
I_s=\int_{r_{+}}^{\infty}~dr~r^{-s}\left(\frac{r}{r_{+}}-1 \right)^{i\omega/\kappa_{+}}
\left(\frac{r}{r_{-}}-1 \right)^{-i\omega/\kappa_{-}}e^{2i\omega
r},\quad s=2,3,4.
\]
By means of the substitution $x=(r/r_{+})-1$ the above integral
becomes up to a constant multiplicative factor
\[
\mathcal{I}_s=\int_{0}^{\infty}~dx~h(x),\quad
h(x)=\frac{x^{i\frac{\omega}{\kappa_{+}}}}{(x+1)^s(x+\Delta 
)^{i\frac{\omega}{\kappa_{-}}}}~e^{2i\omega x},\quad \Delta
=(r_{+}-r_{-})/r_{+} >0.
\]
Taking into account that
\[
|h(x)|\approx\left\{ \begin{array}{ll}
         x^{-\frac{\omega_I}{\kappa_{+}}} & \mbox{as $x\to 0^{+}$},\\
         x^{\omega_I\left(\frac{1}{\kappa_{-}}-\frac{1}{\kappa_{+}}\right)-s}e^{-2\omega_I x} & \mbox{as $x\to\infty$},\end{array}
         \right.,\quad s=2,3,4
\]
the integral $\mathcal{I}_s$ will exist if the imaginary part of
the quasinormal frequency is restricted to the following range
\begin{equation}\label{conditio-RN}
0\leq\omega_I<\kappa_{+}.
\end{equation}
Let $\omega_I\in[0,\kappa_{+})$. Unfortunately the above integral
cannot be computed in closed form. However, it takes a dominant
contribution  close to the event horizon $x\approx 0$. Hence, we
shall consider the integral
\[
\int_{0}^{\infty}~dx~x^{i\frac{\omega}{\kappa_{+}}}e^{2i\omega
x}=\frac{i}{2\omega(-2i\omega)^{\frac{i\omega}{\kappa_{+}}}}~\Gamma\left(1+i\frac{\omega}{\kappa_{+}}\right).
\]
From this result we immediately see that the imaginary parts of
the quasinormal modes will be given by
\[
\omega_n=in\kappa_{+},\quad n\gg 1.
\]
However, these values should be disregarded in view of the
condition (\ref{conditio-RN}).
In the extreme case $M=Q$ the event
and Cauchy horizon coalesce into a single horizon at $r_0=M$.
Then, the spatial function $f(r)$ can be
rewritten as $f(r)=r^2/(r-M)^2$, whereas the tortoise coordinate
is given by
\[
r_{*}=M\left[\rho-\frac{1}{\rho-1}+\ln{(\rho-1)^2}\right],\quad \rho=\frac{r}{M}.
\]
By means of the above relation and the transformation $x=\rho-1$ the scattering amplitude can be
written as
\[
f^{1D}_{Born}(\Omega)=\frac{e^{i\Omega}}{M}\int_{0}^{\infty}dx\left[\frac{\ell(\ell+1)}{(x+1)^2}+\frac{2}{(x+1)^3}-\frac{2Q^2}{M^2(x+1)^4}\right]x^{2i\Omega}e^{i\Omega\left(x-\frac{1}{x}\right)},
\]
where $\Omega=2M\omega$. Essentially we have to solve integrals of the form
\begin{equation}\label{integralozzo}
\mathcal{I}_s=\int_{0}^{\infty}~dx~\frac{x^{2i\Omega}}{(x+1)^s}~e^{i\Omega\left(x-\frac{1}{x}\right)}
\end{equation}
with $s=2,3,4$. Taking into account that
\[
\left|\frac{x^{2i\Omega}}{(x+1)^s}~e^{i\Omega\left(x-\frac{1}{x}\right)} \right|\approx\left\{
\begin{array}{ll}
         x^{-2\Omega_I}e^{\frac{\Omega_I}{x}} & \mbox{as $x\to 0^{+}$}~\Longrightarrow~\Omega_I<0,\\
         x^{-2\Omega_I-s}e^{-\Omega_I x} & \mbox{as $x\to\infty$}~\Longrightarrow~\Omega_I>0,\end{array}
         \right.,\quad s=2,3,4
\]
the integral $\mathcal{I}_s$ has a chance to exist only if $\Omega_I=0$. Let
$\Omega_I=0$ in (\ref{integralozzo}). Since the corresponding
integral cannot be computed in closed form, we shall consider only
the dominant contribution close to the event horizon $x\approx 0$.
Hence, we obtain \cite{GRAD}
\[
\int_{0}^{\infty}~dx~x^{2i\Omega_R}e^{-i\frac{\Omega_R
}{x}}=(i\Omega)^{2i\Omega+1}\Gamma(-1-2i\Omega_R).
\]
From this result we immediately see that the poles of the Gamma
function should occur at the complex values
\[
\Omega_{R,n}=-\frac{in}{2},\quad n\gg 1
\]
which is impossible. In 
this case too, the Born approach is inconclusive for the determination
of the poles of the scattering amplitude.

\subsection{Reissner-Nordstr\"{o}m-de Sitter metric}
In the present case
\[
f(r)=1-\frac{2M}{r}+\frac{Q^2}{r^2}-\frac{\Lambda r^2}{3}
\]
where the constants $M>Q$, $Q\in\mathbb{R}$ are the mass and
electric charge of the black hole and $\Lambda\geq 0$ is the
cosmological constant. In general, the equation $f(r)=0$ has four
different roots: a negative root $r_1$ without a physical meaning, a
minimum positive root $r_c$ corresponding to a Cauchy horizon, a
root $r_{-}>r_c$ interpreted as an event horizon and a root
$r_c>r_{-}>r_c$ corresponding to a cosmological horizon. According
to \cite{KOB,ZHAO} these horizons are given by
\[
r_{+}=-A+B+C,\quad r_{-}=A-B+C,\quad r_c=A+B-C
\]
where
\begin{eqnarray*}
A&=&\frac{1}{\sqrt{2\Lambda}}\sqrt{1-\sqrt{1-4\Lambda Q^2}\cos\left(\frac{\alpha}{3}-\frac{\pi}{3}\right)},\\
B&=&\frac{1}{\sqrt{2\Lambda}}\sqrt{1-\sqrt{1-4\Lambda Q^2}\cos\left(\frac{\alpha}{3}+\frac{\pi}{3}\right)},\\
C&=&\frac{1}{\sqrt{2\Lambda}}\sqrt{1+\sqrt{1-4\Lambda
Q^2}\cos\left(\frac{\alpha}{3}\right)},
\end{eqnarray*}
and
\[
\alpha=\arccos{\left[-\frac{1-6\Lambda(3M^2-2Q^2)}{(1-4\Lambda Q^2
)^{3/2}}\right]}.
\]
In the exterior region $r_{-}<r<r_{+}$ we introduce a tortoise
coordinate $r_{*}$ by means of the relation
\[
\frac{dr_{*}}{dr}=\frac{1}{f(r)}.
\]
Taking into account that the surface gravities at the the three
horizons are given by \cite{ZHAO1}
\begin{eqnarray*}
\kappa_{+}&=&\frac{\Lambda}{6r_{+}^{2}}~(r_{+}-r_1)(r_{+}-r_c)(r_{+}-r_{-}),\\
\kappa_{-}&=&\frac{\Lambda}{6r_{-}^{2}}~(r_{-}-r_1)(r_{-}-r_c)(r_{+}-r_{-}),\\
\kappa_{c}&=&\frac{\Lambda}{6r_{c}^{2}}~(r_{c}-r_1)(r_{-}-r_c)(r_{+}-r_{c}),
\end{eqnarray*}
the tortoise coordinate is explicitly given by
\[
r_{*}=\frac{1}{2\kappa_{-}}\ln{\left|\frac{r}{r_{-}}-1\right|}-\frac{1}{2\kappa_{+}}\ln{\left|1-\frac{r}{r_{+}}\right|}
-\frac{1}{2\kappa_c}\ln{\left|\frac{r}{r_c}-1\right|}
+\frac{1}{a}\ln{\left|\frac{r}{r_1}-1\right|}
\]
where
\[
a=\frac{\Lambda}{6r_{1}^{2}}~(r_{c}-r_1)(r_{-}-r_1)(r_{+}-r_{1}).
\]
Hence, the scattering amplitude is given by
\[
f^{1D}_{Born}(\omega)=\int_{r_{-}}^{r_{+}}dr~U(r)
\left(\frac{r}{r_{-}}-1\right)^{i\frac{\omega}{\kappa_{-}}}\left(1-\frac{r}{r_{+}}\right)^{-i\frac{\omega}{\kappa_{+}}}
\left(\frac{r}{r_c}-1\right)^{-i\frac{\omega}{\kappa_{c}}}\left(\frac{r}{r_1}-1\right)^{2i\omega
a }
\]
with
\[
U(r)=\frac{\ell(\ell+1)}{r^2}+\frac{2M}{r^3}-\frac{2Q^2}{r^4}-\frac{2}{3}\Lambda.
\]
By means of the transformation $u=(r-r_{-})/(r-r_{+})$ mapping the
event horizon to $0$ and the cosmological horizon to $1$ we can
reduce the computation of $f^{1D}_{Born}(\omega)$ to the computation of the
following integral
\[
\mathcal{I}_s=\int_0^1 du G_s(u),\quad s=0,2,3,4
\]
where
\[
G_s(u)=\left\{
\begin{array}{ll}
          \frac{u^{i\frac{\omega}{\kappa_{-}}}(1-u)^{-i\frac{\omega}{\kappa_{+}}}}{(1-yu)^s(1-\Xi u)^{i\frac{\omega}{\kappa_c}}(1-wu)^{2i\omega a}} & \mbox{for $s=2,3,4$},\\
          \frac{u^{i\frac{\omega}{\kappa_{-}}}(1-u)^{-i\frac{\omega}{\kappa_{+}}}}{(1-\Xi u)^{i\frac{\omega}{\kappa_c}}(1-wu)^{2i\omega a}} & \mbox{for $s=0$},\end{array}
         \right.
\]
with $y$ 
\[
\Xi=\frac{r_{-}-r_{+}}{r_{-}-r_c},\quad
w=\frac{r_{-}-r_{+}}{r_{-}-r_{1}}, \quad
y=\frac{r_{-} - r_{+}}{r_{-}}.
\]
Taking into account that
\[
|G_{s}(u)|\approx\left\{
\begin{array}{ll}
         u^{-\frac{\omega_I}{\kappa_{-}}} & \mbox{as $u\to 0^{+}$}~\Longrightarrow~\omega_I<\kappa_{-},\\
         (1-u)^{\frac{\omega_I}{\kappa_{+}}} & \mbox{as $u\to 1^{-}$}~\Longrightarrow~\omega_I>-\kappa_{+},\end{array}
         \right.,\quad s=0,2,3,4
\]
it follows that the above integral exists if and only if the
imaginary part of the quasinormal mode satisfies the condition
\begin{equation}\label{cond-RN-Sds}
-\kappa_{+}<\omega_I<\kappa_{-}.
\end{equation}
Fortunately, the integral $\mathcal{I}_s$ can be computed in
closed form in terms of Gamma functions and the Lauricella
function $F^{(n)}_D$ with $n=3$ whose Picard's integral
representation is given by \cite{PIC,TAN}
\[
\int_0^1 du~u^{a-1}(1-u)^{c-a-1}\prod_{i=1}^{n}(1-x_i
u)^{-b_i}=\frac{\Gamma(a)\Gamma(c-a)}{\Gamma(c)}~F^{(n)}_D(a,b_1,b_2,\cdots,b_n;c;x_1,x_2,\cdots,x_n)
\]
with
\[
F^{(n)}_D(a,b_1,b_2,\cdots,b_n;c;x_1,x_2,\cdots,x_n)=
\sum_{i_1,i_2,\cdots,i_n=0}^{\infty}\frac{(a)_{i_1+i_2+\cdots+i_n}(b_2)_{i_2}\cdots(b_n)_{i_n}}
{(c)_{i_1+i_2+\cdots+i_n}i_1 ! i_2 !\cdots i_n
!}x_1^{i_1}x_2^{i_2}\cdots x_n^{i_n}
\]
absolutely convergent for $\max{\{|x_1|,|x_2|,\cdots,|x_n|\}}<1$.
An analysis similar to that performed in \cite{CHO} shows that the
poles of the scattering amplitude are located at
\[
\omega_n=in\kappa_{-},\quad\omega_n=-in\kappa_{+}\quad n\gg 1
\]
but we should disregard them in virtue of the integrability
condition (\ref{cond-RN-Sds}).

\section{Quasinormal modes from a Coulomb-like phase shift}
For large $r$ one can expect that the scattering can be described by a
$1/r$ Newtonian potential. Indeed, in \cite{kofinti}, such an analysis has
been explicitly performed. Starting with the Klein-Gordon equation,
$\square \Psi = g_{\mu \nu} \nabla^{\mu} \nabla^{\nu} \Psi = 0$ and
using the separation ansatz, 
\begin{equation}
\Psi(t,r,\theta,\phi) = T(t) R(r) \Theta(\theta, 
\phi) \,, 
\end{equation}
the transformation 
\begin{equation}
R = u(r) [r (1 - 2M/r)]^{-1/2}
\end{equation} 
reveals the asymptotic form,
\begin{equation}
R\, {\buildrel r \to \infty \over \longrightarrow }\,
(1/r) \, \sin[\omega r - \gamma\ln (2\omega r) - (1/2)l \pi - 
\sigma_l]\,, 
\end{equation} 
where $\gamma = - 2 M \omega$ and $\sigma_l$ is the equivalent
Coulomb phase shift given by
\begin{equation} 
\sigma_l = {\rm arg} \Gamma (l + 1 - 2iM\omega).  
\end{equation}
Using the property $\Gamma(\bar{z}) = \overline{\Gamma(z)}$, the partial wave
scattering matrix is given by
\begin{equation} 
e^{2 i \sigma_l} = {\Gamma (l + 1 - 2iM\omega) \over 
\Gamma (l + 1 + 2iM\omega)}.
\end{equation}
The poles corresponding to the above amplitude are
$\omega_n = - i \kappa\, 2(n + l +1)$ which for large $n$ can be written as
$\omega_m = - i \kappa\,m$. It therefore appears that the large imaginary
parts of the QNM frequencies are nothing but the poles of the Coulomb-like
amplitude.

Two comments are in order. Let us first note that in \cite{kofinti} the 
signature of the metric is $(1,-1,-1,-1)$ and therefore the time development 
is proportional to $\exp(-i \omega t)$. With $\omega = - i \kappa m$, this 
corresponds to a decaying state $\exp(-\kappa m t)$. Physically this case is 
equivalent to working with a signature $(-1,1,1,1)$ and the time 
evolution $\exp(+i \omega t)$ with the QNMs given by $\omega = + i \kappa m$. 
Secondly we remark that in atomic physics the definition of the principal 
quantum number is $n = j_{max} + l +1$ which is very similar to the 
definition of $m$ above. 

\section{Variable amplitude method for black hole scattering} 
The Schr\"odinger like equation (\ref{radial}) in black hole scattering 
can be solved using standard techniques for tunneling in quantum mechanics. 
The asymptotic solutions of the Schr\"odinger 
equation (\ref{radial}) are: 
$$\psi \, =\, A(\omega) e^{+i\omega r_*} \, +\, B(\omega) 
e^{-i\omega r_*}, \,\,\,\, r_* \to - \infty$$
$$\psi \, =\, C(\omega) e^{+i\omega r_*} \, +\, D(\omega) e^{-i\omega r_*}, 
\,\,\,\, r_* \to + \infty .$$
For waves incident on the black holes from the right (i.e. $+\infty$), 
$B(\omega)=0$, the reflection amplitude  
$R(\omega) = D(\omega)/C(\omega)$
and the transmission amplitude 
$T(\omega) = A(\omega)/C(\omega)$, so that
$$\psi \, =\,  T(\omega) e^{i\omega r_*}, \,\,\,\, r_* \to - \infty$$
$$\psi \, =\, e^{i\omega r_*} \, +\, R(\omega)  e^{-i\omega r_*}, 
\,\,\,\, r_* \to + \infty. $$
The reflection amplitude $R(\omega)$ is related to the scattering amplitude 
$f^{1D}(\omega)$ in one dimension (1D) as 
$R(\omega) = if^{1D}(\omega)/2\omega$, where 
\begin{equation}\label{fexact}
f^{1D}(\omega) =  \int_{-\infty}^{\infty} \, 
e^{i \omega x} \, V(x)\, \psi(x) dx  
\end{equation}
and reduces to (\ref{born1D}) in the Born approximation. We shall evaluate 
$R(\omega)$ numerically via the variable amplitude method. This $R(\omega)$ 
corresponds to the exact scattering amplitude $f^{1D}(\omega)$ 
and will be compared with $R^{Born}(\omega) = if^{1D}_{Born}(\omega)/2\omega$. 
The variable amplitude method was first introduced in \cite{tikochin} and 
has been widely used to evaluate the reflection and transmission coefficients 
for different potentials in literature \cite{rozkidunlee}. 

The scattering matrix in 1D is given as 
\begin{equation}\label{4}
{\bf S} \,=\,\left(
\begin{array}{cc}
T_L(\omega) & R_R(\omega) \\ 
R_L(\omega) & T_R(\omega)
\end{array}
\right )
\end{equation}
where $R_L(\omega)$, $T_L(\omega)$ 
are the reflection and transmission amplitudes respectively 
for incidence from the left ($- \infty$) and $R_R(\omega)$, $T_R(\omega)$ 
for incidence from the right ($+\infty$).
For elastic scattering, $T_L(\omega) = T_R(\omega)$. 
Though the amplitudes for left and right 
incidence may not necessarily be equal, the reflection and transmission 
coefficients are the same, i.e., $|R_L(\omega)|^2 = |R_R(\omega)|^2 = 
\cal{R}(\omega)$ and $|T_L(\omega)|^2 = |T_R(\omega)|^2 = \cal{T}(\omega)$.
Starting with incidence from the right (and dropping the subscript $R$ 
for convenience), we 
follow the standard procedure \cite{rozkidunlee} to obtain the 
Riccati equation for black hole scattering. 
This involves writing the solution of the Schr\"odinger equation as
a superposition of the reflected and transmitted waves, namely,  
$\psi(\omega,r_*) = T(\omega,r_*) [e^{i\omega r_*} + R(\omega,r_*) 
e^{-i\omega r_*}]$, which 
leads to the following equation for $R$:
\begin{equation}\label{Riccati1}
\frac{dR(\omega,r_*)}{dr_{*}}=-\frac{V(r_{*})}{2i\omega}\left[e^{i\omega
r_{*}}+R(\omega,r_{*})e^{-i\omega r_{*}}\right]^2 . 
\end{equation}
The absence of reflection behind the potential at $r_* \to - \infty$ imposes 
the boundary condition $R(\omega, -\infty) = 0$ on the above equation. 
The reflection 
coefficient ${\cal R}(\omega)$ is given by ${\cal R} (\omega) = 
|R(\omega,\infty)|^2$. 
Solving the above equation (\ref{Riccati1}) can be made easier by 
introducing a new function $U(\omega,r_*)$ such that 
\begin{equation}\label{requation}
R(\omega,r_*) \, =\, exp(-2 i \omega r_*) \, 
[ 2 i \omega U(\omega,r_*)\, -\, 1] . 
\end{equation}
The equation for $U$ is,
\begin{equation}\label{uequation}
{dU(\omega,r_*) \over dr_*} \, =\, 1\, -\, 2 i \omega U(\omega,r_*)\, -
\, V(r_*)\, U^2(\omega,r_*). 
\end{equation}
Eq. (\ref{uequation}) is solved numerically with the boundary 
condition $U(\omega,-\infty)=1/2i\omega$ 
to determine ${\cal R} (\omega)= |R(\omega, \infty)|^2$. 
\begin{figure}[h]
\includegraphics[width=11cm,height=9cm]{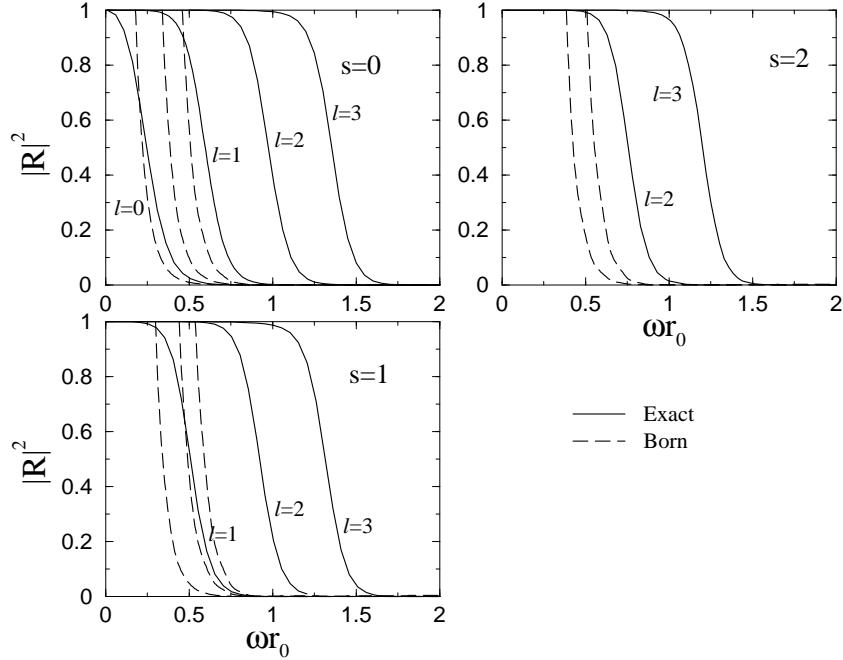}
\caption{\label{fig:eps1}
Reflection coefficient or the scattering amplitude squared for the 
scattering of massless scalar ($s = 0$), electromagnetic ($s = 1$) and 
gravitational ($s =2$) waves from a black hole.
The solid lines give the exact results for different $l$. The dashed 
lines represent the Born approximation results and follow the same 
sequence for $l$ from left to right.}
\end{figure} 

In Fig. 1 we compare the reflection coefficient obtained by numerically 
solving the Riccati equation with that obtained within the Born 
approximation, namely, $|R^{Born}(\omega)|^2 = 
|if^{1D}_{Born}(\omega)/2\omega|^2$. 
The formalism of section II can be generalized in the Schwarzschild case 
to arbitrary spin $s$ of the scattered particle. The Regge-Wheeler potential 
entering the Schr\"odinger equation is
\begin{equation}
V(r_*) = \biggl (1 - {2M \over r}\biggr)\, \biggl[ {l(l+1)\over r^2} 
+ {2M (1 - s^2)\over r^3} \biggr ]
\end{equation}
where, $l \ge s$. 
The Born approximation in general is 
not expected to be very useful at low energies. However, we find from Fig. 1 
that it does not agree even qualitatively with the exact result for 
any of the three cases considered. One can also see that the reflection 
coefficient at large energies is very small and any agreement 
(if at all) with the 
Born amplitude at high energies becomes irrelevant. 

\section{Summary}
The Born approximation for particle scattering from black holes has been 
studied in detail for the most relevant black hole spacetimes. 
In summary we can say:
\begin{enumerate}
\item We succeeded in writing the integral of the scattering amplitude 
in the Born approximation in terms of special functions. 
The quasinormal modes can in principle be extracted from this 
parametrization. 
\item The conditions for the existence of the Born integral impose a 
validity range on $\omega$. In all cases of black hole spacetimes we 
find that the quasinormal modes found here as well as 
elsewhere in literature using the Born approximation lie outside the 
validity range. Therefore, it is mathematically impossible to find 
the quasinormal modes within the Born approximation.
\item We have shown that the quasinormal modes with a large imaginary 
part can be obtained from the Coulomb-like phase shift in black hole 
scattering.
\item Comparing the reflection coefficient within the Born approximation 
with the exact one evaluated numerically (for the case of 
Schwarzschild black hole) 
reveals that the only reasonable 
agreement occurs for the $s=0$, $l=0$ case.
\end{enumerate}

\end{document}